%% file: NM_emulator_PRL.tex
\documentclass[aps,prl,twocolumn,superscriptaddress,longbibliography]{revtex4-1}
\usepackage{epsfig,dsfont,amssymb,amsmath,amsthm,amsfonts,amsbsy,mathrsfs}
\usepackage{graphicx}
\usepackage{amsmath}
\usepackage{amssymb}%
\usepackage{dcolumn}
\usepackage{hyperref}
\usepackage[color=green!30]{todonotes}

\newcommand{\be}{\begin{equation}}
\newcommand{\ee}{\end{equation}}
\newcommand{\ba}{\begin{eqnarray}}
\newcommand{\ea}{\end{eqnarray}}

\input{./shorthands.tex}

\begin{document}

\title{Nuclear-matter saturation and symmetry energy within $\Delta$--full chiral effective field theory}

\author{W.\ G.\ Jiang}
\affiliation{Department of Physics, Chalmers University of Technology, SE-412 96 G\"oteborg, Sweden}
\affiliation{Institut f\"ur Kernphysik and PRISMA Cluster of Excellence, Johannes Gutenberg Universit\"at, 55128 Mainz, Germany}

\author{C.\ Forss\'en}
\affiliation{Department of Physics, Chalmers University of Technology, SE-412 96 G\"oteborg, Sweden}

\author{T.\ Dj\"arv}
\affiliation{Department of Physics, Chalmers University of Technology, SE-412 96 G\"oteborg, Sweden}
\affiliation{Physics Division, Oak Ridge National Laboratory, Oak Ridge, TN 37831, USA}

\author{G.\ Hagen}
\affiliation{Physics Division, Oak Ridge National Laboratory, Oak Ridge, TN 37831, USA}
\affiliation{Department of Physics and Astronomy, University of Tennessee, Knoxville, TN 37996, USA}

\begin{abstract}
  Nuclear saturation and the symmetry energy are key properties of low-energy nuclear physics that depend on fine details of the nuclear interaction. The equation-of-state around saturation is also an important anchor for extrapolations to higher densities and studies of neutron stars.
  Here we develop a unified statistical framework that uses realistic nuclear forces to link the theoretical modeling of finite nuclei and infinite nuclear matter. We construct fast and accurate emulators for nuclear-matter observables and employ an iterative history-matching approach to explore and reduce the enormous parameter domain of $\Delta$-full chiral interactions.
  We perform rigorous uncertainty quantification and find that model calibration including \nuc{16}{O} observables gives saturation predictions that are more precise than those that only use few-body data.
\end{abstract}

\maketitle
{\it Introduction---}%
A key question in low-energy nuclear physics is whether it is possible to successfully describe all systems from finite nuclei to infinite nuclear matter using nucleons as effective degrees of freedom.
Realistic interaction models based on \chiEFT{} used in combination with \emph{ab initio} methods, that solve the many-body problem with controlled approximations, have the potential to deliver on that research program~\cite{machleidt2011,hebeler2011,ekstrom2013,ekstrom2015a,jiang2020,soma2020,Hu:2021trw,maris2022,elhatisari2022}. However, using a complicated interaction model and calibration data with limited independence implies a risk for overfitting unless relevant theoretical uncertainties are accounted for. Furthermore, the extrapolation from well-studied few-nucleon systems to heavier nuclei and infinite nuclear matter will lead to increasing variances in model predictions.
Therefore it becomes important to study the precision of these predictions and to explore the sensitivity to different choices of calibration data.
The emergence of nuclear saturation---represented by a minimum in the \EOS{} for infinite \SNM{}---is particularly important since it affects bulk properties (such as binding energies and radii) of finite atomic nuclei. Furthermore, the density dependence of the \EOS{} for \PNM{} is central for the physics of neutron stars \cite{Dietrich:2020efo,Huth:2021bsp,Lattimer:2023rpe}.

In the last decade significant progress has been made towards quantifying EFT uncertainties of \emph{ab initio} nuclear matter predictions and identifying possible correlations with observables in finite nuclei~\cite{hebeler2011,khan2012,hagen2015,kievsky2018,drischler2019,reinhard2016, carson2018, drischler2020a, drischler2020b, Hu:2021trw}. Very recently, some of the authors of this work identified correlations between the symmetry energy and its slope with the neutron skin and dipole polarizability of the heavy nucleus $^{208}$Pb starting from chiral interactions at \NNLO{} with explicit delta isobars ($\Delta$)~\cite{Hu:2021trw}. This finding was made possible by employing novel statistical tools such as bayesian inference~\cite{udo2011} and history matching~\cite{vernon2010,vernon2014,vernon2018,Hu:2021trw} together with accurate emulators of \emph{ab initio} computations of light nuclei~\cite{ekstrom2019,Konig:2019adq,wesolowski2021,Djarv:2021pjc}. 

In this Letter we develop fast and accurate emulators of coupled-cluster computations of \PNM{} and \SNM{} ~\cite{baardsen2013,hagen2013b} starting from $\Delta$-full \chiEFT{} interaction models at \NNLO{}~\cite{vankolck1994,hemmert1998,kaiser1998,krebs2007,epelbaum2008,ekstrom2017,jiang2020}. Our construction of emulators is based on the \SPCC{} method~\cite{ekstrom2019}, which we here extend with small batch voting~\cite{jiang2022:long}---a new on-the-fly validation approach that addresses the possible appearance of spurious states when diagonalizing a Hamiltonian in a subspace of bi-orthogonal coupled-cluster solutions~\cite{shavittbartlett2009}. Using these emulators we can reproduce full-space coupled-cluster computations of infinite nuclear matter with high precision at a tiny fraction of the computational cost.

After validating our approach, we use history matching to identify a region of the 17-dimensional parameter space of the \LEC[s] at \NNLO{} that give acceptable results when confronted with few-body data. This iterative parameter search is enabled by employing emulators of few-body systems up to {\nuc{4}{He}} and allows us to collect $1.7 \times 10^6$ non-implausible interaction samples.  
Finally, we calibrate our \emph{ab initio} model with two alternative \LEC{} posterior \PDF[s] and use the principle of importance resampling~\cite{Smith:1992aa,Jiang:2022off} to quantify the sensitivity of nuclear matter predictions to calibration data.
Here we focus on the results of this analysis, and the emergence of nuclear saturation, while details of the small-batch voting scheme and the history match are presented in a companion paper~\cite{jiang2022:long}.
  
{\it Method---}%
The chiral Hamiltonian of $\Delta$\NNLO{} is parametrized with 17 \LEC[s]{}~\cite{jiang2020}, and following Refs.~\cite{ekstrom2019,Konig:2019adq,wesolowski2021} it can be written as
\begin{eqnarray}
  \label{eq_H_LECs}
  H(\alpha) = h_{0} +  \displaystyle\sum_{i=1}^{N_{\rm{LECs}}=17}  \alpha_i h_i.
\end{eqnarray}
Here, $h_0 = t_{\rm{kin}}+v_0$ where $t_{\rm{kin}}$ is the kinetic energy and $v_0$ represents the constant potential term without \LEC{}-dependence. Note that ${\vec \alpha}$ is a vector that denotes all \LEC[s].
In this work we use non-local regulators with a cutoff $\Lambda = 394$~MeV/$c$. 

Recently, model reduction methods~\cite{melendez2022} such as \EC{}~\cite{Frame2018,sarkar2020}, has proven to be both efficient and accurate for emulating the output of \emph{ab initio} computations of both scattering~\cite{melendez2021, Zhang:2021jmi} and bound-state observables~\cite{ekstrom2019,Konig:2019adq,wesolowski2021,Djarv:2021pjc,Demol:2019yjt}.
These methods employ the fact that the eigenvector trajectory generated by smooth changes of the Hamiltonian matrix is well approximated by a very low-dimensional manifold~\cite{Frame2018}. Following Refs.~\cite{ekstrom2019,Konig:2019adq} we can therefore obtain good approximations of the ground-state of a target Hamiltonian $H({\vec \alpha}_{\circledcirc})$ by projecting it on a subspace of $N_{\rm{sub}}$ different ground-state training eigenvectors and solving the corresponding $N_{\rm{sub}} \times N_{\rm{sub}}$-dimensional generalized eigenvalue problem.
In this work we employ emulators for \nuc{2,3}{H} and \nuc{4}{He} for history matching, and the \SPCC{} method with singles-, doubles-, and leading order triples excitations (CCSDT-3) for subsequent model calibration on $^{16}$O (see Ref. ~\cite{Hu:2021trw} for details). We also construct a new emulator for \nuc{6}{Li} based on Ref.~\cite{Djarv:2021pjc} for model checking. Here we used $N_\mathrm{sub} = 32$ training points reaching a relative accuracy of $10^{-3}$ for the non-implausible parametrizations~\cite{djarv:2021phd}. 

In this Letter we use the coupled-cluster method~\cite{coester1958,coester1960,cizek1966,kuemmel1978,bishop1991,zeng1998,mihaila2000b,dean2004,bartlett2007,shavittbartlett2009,hagen2010b,binder2013,hagen2014} within the doubles 
approximation (CCD) and solve for pure neutron and symmetric nuclear matter on a cubic lattice in momentum space using periodic boundary conditions~\cite{hagen2013b}. The model-space has $(2n_{\rm{max}}+1)^3$ momentum points with $n_{\rm{max}}=4$. We use 132 nucleons for \SNM{} and 66 neutrons for \PNM{} which allows to minimize finite-size effects~\cite{gandolfi2009,hagen2013b}. 
We then use the \EC{}-inspired \SPCC{} method~\cite{ekstrom2019} to construct subspace emulators at five different nuclear densities. 
Since the subspace projected coupled-cluster Hamiltionian is non-Hermitian, the variational theorem does not hold~\cite{teschl2009}. This might lead to the appearance of ``spurious'' states in the subspace spectrum that are lower in energy than the corresponding full-space coupled-cluster solution for certain combinations of the \LEC[s]. These states appear due to the non-symmetric treatment of the left and right bi-orthogonal coupled-cluster states~\cite{bartlett2007}.  
Therefore, we have developed a new  algorithm---called {small-batch voting}---which allows to identify the physical ground state using a set of different subspace projections. This algorithm employs the strong sensitivity of spurious states on the specific choice of basis, and the contrasting stability of physical states. Details of this algorithm are given in the companion paper~\cite{jiang2022:long}.

{\it Results---}%
We used  history matching to collect a large number of $1.7 \times 10^6$ model parametrizations that exhibit acceptable (or at least non-implausible) fits to the history-matching observables.
The latter include a total of 36 neutron-proton phase shifts in $S$ and $P$ waves up to $T_\mathrm{lab}=200$~MeV scattering energy and six bound-state observables for $A=2,3,4$ systems. This set is relevant for model calibration, allows fast model simulation or emulation, and permits the construction of simple implausibility measures.
In the final wave we defined the non-implausible volume using a rotated hyperrectangle that captured parameter correlations and allowed to significantly increase the number of collected non-implausible samples. Further details of the history match are presented in the companion paper~\cite{jiang2022:long}. We then strategically selected 64 of the most accurate non-implausible samples for which we performed full CCD computations for \SNM{} and \PNM{} at five different densities $\rho \in \{ 0.12, 0.14, 0.16, 0.18, 0.20\}$~fm$^{-3}$. Together with small-batch voting this allowed to create \SPCC{} emulators for $E/A$ and $E/N$ at these densities. Finally, the nuclear matter \EOS{} around saturation could be obtained by interpolation using Gaussian processes~\cite{rasmussen2006} as described in more detail in Ref.~\cite{jiang2022:long}. 
 
\begin{figure}
  \includegraphics[width=0.95\columnwidth] {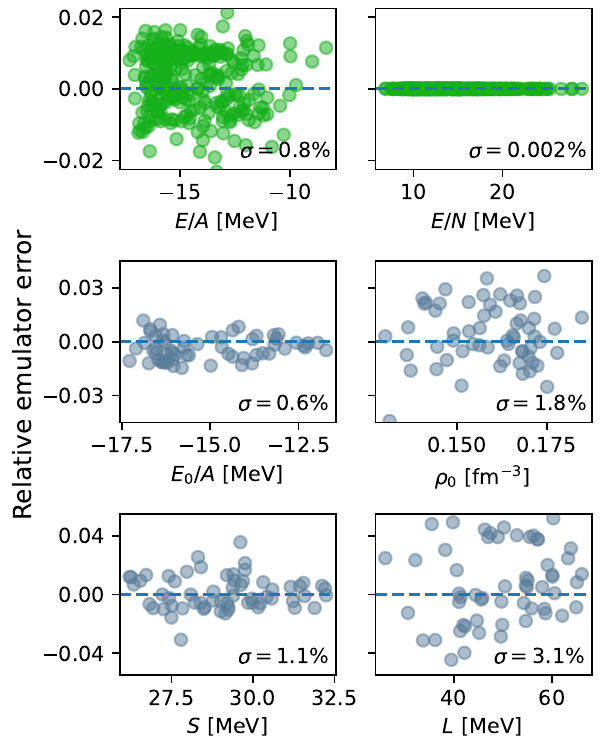}
  \caption{(Color online) 
   Cross-validation of the \SPCC{} emulator with exact CCD calculations using 50 interaction samples. Shown here from top to bottom is the nuclear matter saturation density $(\rho_0)$, saturation energy ($E_0/A)$, and symmetry energy $(S)$. Histograms of relative errors are shown in the right column.
   \label{fig:cross-validation}%
 }
\end{figure}

Cross validation of emulator performance is shown in Fig.~\ref{fig:cross-validation}. The 50 validation interactions were randomly selected from the non-implausible volume in a space-filling manner. We conclude that our emulators predict the energy per particle for SNM (PNM) with $< 10^{-2}$ ($< 10^{-4}$) relative precision at a computational cost that is six (eight) orders of magnitude smaller than the full solution. Furthermore, Gaussian-process interpolation of the emulated EOS allows to extract empirical nuclear-matter properties with $\approx 1\%$ precision (increasing to 3\% and 10\% for derivative quantities $L$ and $K$, respectively). 

\begin{figure*}[htbp]
  \includegraphics[width=0.90\textwidth] {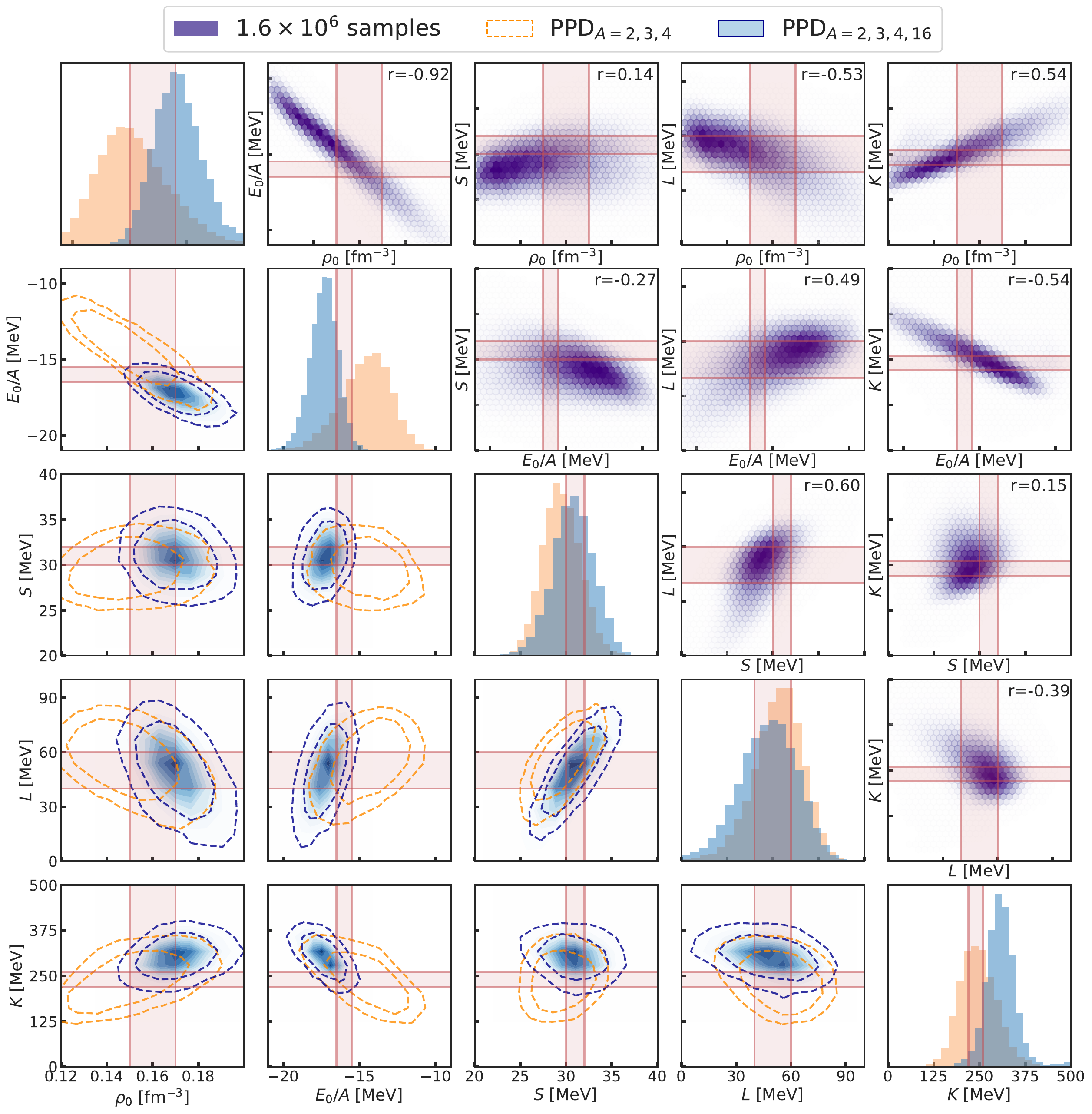}
  \caption{(Color online) Upper triangle: nuclear matter emulator output (saturation density $\rho_0$, saturation energy $E_{0}/A$, symmetry energy $S$, symmetry energy slope $L$ and incompressibility $K$) for the non-implausible interactions from the fifth wave of history matching. The axes limits are the same as in the corresponding panels in the lower triangle.
    Lower triangle and diagonal: \PPD[s] for nuclear matter properties using two different \PDF[s] for the \LEC[s]{} plus error sampling. These predictions are based either on few-body ($A=2,3,4$) calibration data (orange \PPD{}) or the addition of \nuc{16}{O} to the calibration data set (blue \PPD{}).
    See the Supplemental Material~\cite{supp2022:PRL} for a Gaussian approximation of \ppdmany{}.
    The contour lines on the bivariate distributions denote $68\%$ and $90\%$ credible regions.
    The red bands indicate empirical ranges for $E_0/A = -16.0 \pm 0.5$~MeV, $\rho_0 = 0.16 \pm 0.01$~fm$^{-3}$,  $S=31 \pm 1$, $L=50\pm 10$ and $K=240\pm20$ ~MeV from Refs.~\cite{lattimer2013,bender2003, shlomo2006}.
  \label{fig:NM_corner_plot}}
\end{figure*}

Having access to fast and accurate emulators, and having identified a non-implausible region in the parameter space of our chiral interaction model, we can proceed to study the general behaviour of nuclear matter model predictions and possible correlations between different properties of these systems. Here we are not interested in the usual optimization approach that results in a single (``best fit'') model parametrization. Instead, we consider all non-implausible samples from the history match. About 73\% of these samples predict saturation within the studied density region and are kept for further consideration. 
The outputs from the nuclear matter emulators for these $ 1.2 \times 10^6$ interaction samples are shown in the upper triangle of Fig.~\ref{fig:NM_corner_plot}. Here we have applied a density-dependent energy shift to approximately account for triples corrections~\cite{jiang2022:long}.

We observe a strong anti-correlation between the saturation energy $E_0/A$ and the saturation density $\rho_0$ (Pearson correlation coefficient $r=-0.92$). This finding is in agreement with the Coester line~\cite{Coester:1970ai} for nucleon-nucleon interactions. Similarly, the symmetry energy $S$ and its slope $L$ show a positive correlation ($r=0.60$). These correlations have been seen in DFT calculations~\cite{Ducoin:2011fy,Santos:2014vda} and have been indicated with small families of EFT-inspired~\cite{kievsky2018} or ($\Delta$-less) chiral interactions~\cite{drischler2019}. The comparison with empirical nuclear matter properties reveals that the model predictions from the history match are clustered in a region with too small $|E_0/A|$, $\rho_0$, and $S$. This finding is consistent with previous results from various few-body optimized interactions~\cite{Coraggio:2014nva,ekstrom2017,drischler2020b}. 

The non-implausible predictions of nuclear matter properties that is shown in the upper triangle of Fig.~\ref{fig:NM_corner_plot} is a form of Bayes linear forecasting. By just considering expectation values such as means and variances---thereby avoiding the full probabilistic specification of uncertain quantities---we made the analysis simpler and more technically straightforward. While this allowed us to identify the parameter region of interest and to explore the model's forecasting capabilities, we now make an effort to extract a \PPD{} with which we can make probabilistic statements.

First, we revisit all remaining interaction parametrizations and keep only those that survive an extra implausibility analysis involving the selected $S$ and $P$ wave phase shifts and that give an unbound ${}^1S_0$ ground-state in the neutron-proton system. This step results in a significant reduction to $8218$ acceptable samples.
Second, we introduce a set of calibration observables \Dcal{} and a normally-distributed likelihood,
\likelihood{\Dcal}{{\vec \alpha}}, assuming independent experimental, method, emulator, and model errors~\cite{jiang2022:long}.
At this stage we employ the established method of sampling/importance resampling~\cite{Smith:1992aa,Jiang:2022off} to approximately extract samples from the parameter posterior \prCond{{\vec \alpha}}{\Dcal,I}. In this step we assume a uniform prior probability for all non-implausible samples~\footnote{The prior for $c_{1,2,3,4}$ is the multivariate Gaussian resulting from a Roy-Steiner analysis of $\pi N$ scattering data~\cite{siemens2017}.}.

In this work we especially considered two different sets of model calibration data: (i) $\Dcal = \Dfew$ encompassing binding energies and point-proton radii of \nuc{2,3}{H} and \nuc{4}{He} plus the quadrupole moment of the deuteron, and (ii) $\Dcal = \Dmany$ where we complemented the above set of observables with the energy and radius of \nuc{16}{O}. Note that our choice of history-matching observables allows for both PPD resampling analyses to be performed from the same set of non-implausible prior samples.

Having access to approximate parameter posteriors we first extract samples from the model \PPD{} defined as ${\rm PPD}_\mathrm{th} = \{ {\rm \textbf{y}_{th}}({{\vec \alpha}}) : {{\vec \alpha}} \sim \prCond{{\vec \alpha}}{\Dcal,I}  \}$.
As a non-trivial model validation we predict the \nuc{6}{Li} threshold energy $S_d = E(\nuc{2}{H}) + E(\nuc{4}{He}) - E(\nuc{6}{Li})$ using our \EC{} emulators. The 90\% credible intervals are $[-0.14,1.54]$~MeV and $[1.23, 1.84]$~MeV for ${\rm PPD}_\mathrm{th}$ with \Dfew{} and \Dmany{} calibration data, respectively. Both predictions are consistent with the experimental value $S_d = 1.474$~MeV~\cite{Tilley:2002vg}, but the latter one is significantly more precise.

We then predict nuclear matter properties. For these predictions we collect samples from the full \PPD{}, which includes the EFT truncation error, CC method errors, and the \SPCC{} emulator error (estimated from cross validation shown in Fig.~\ref{fig:cross-validation}). The density dependence and relevant cross correlation of these errors are described by a Bayesian machine learning error model~\cite{drischler2020a,drischler2020b,jiang2022:long}. Our strategic emulator construction---using training samples with high importance weight---implies that up to 60\% of resampled interactions have zero emulator error.
Using the two different sets of calibration data, \Dfew{} and \Dmany{}, we compare the resulting \PPD[s]{} that we label \ppdfew{} and \ppdmany{}, respectively.
These are shown in the lower triangle of Fig.~\ref{fig:NM_corner_plot}.

The marginal distributions on the diagonal reveal that $\rho_0$, $E_0/A$ and $S$ for \ppdfew{} are characterized by low precision and mild tension with the empirical region---as previously observed with the non-importance-weighted samples. In other words, even by enforcing a higher accuracy for two- and few-body systems via the data likelihood, the general description of nuclear matter properties is not improved. 
On the other hand, the predictions become more precise and accurate when we include the \nuc{16}{O} observables. In particular, the saturation point is more precisely predicted while its mode shifts to larger saturation density and binding energy.
We actually find that \ppdfew{} displays a significant asymmetry in some dimensions, e.g., $\rho_0$, which hints to a possible bimodality.
Samples from the tail region of \ppdfew{} correspond largely to the mode of \ppdmany{}. This finding shows that the emergence of saturation, represented by the position and shape of this mode, depends on the choice of calibration observables.

\begin{figure}[htbp]
  \includegraphics[width=0.95\columnwidth] {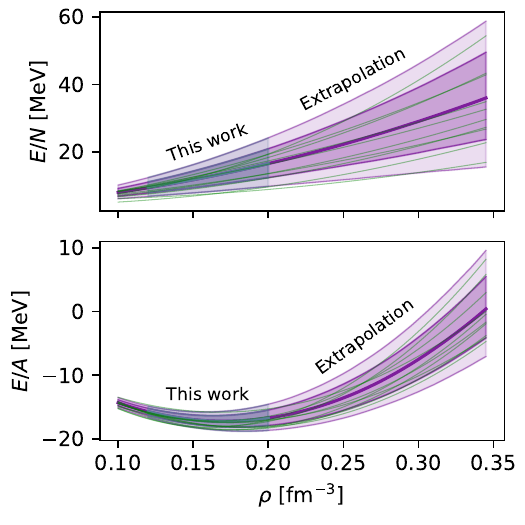}
  \caption{(Color online) Dark (light) blue bands show the \ppdmany{} for the energy per particle of PNM (upper panel) and SNM (lower panel) computed in the $\rho \in [0.12, 0.20]$ fm$^{-3}$ range. The corresponding purple bands represent simple extrapolations based on the empirical nuclear-matter parameters. For each sample of the \PPD{} the \EOS{} is generated by using the expansion of Ref.~\cite{Piekarewicz:2008nh} and retaining up to quadratic terms in $x = (\rho - \rho_0)/3\rho_0$. Ten such samples are shown with thin, gray lines. The extrapolation uncertainty is not incorporated in the error bands.
  \label{fig:NM_EOS}}
\end{figure}

Our predictions of the nuclear \EOS{} around saturation, with quantification of relevant sources of uncertainty, can serve as an important anchor for extrapolations to higher densities and studies of neutron star physics. To illustrate, we show a simple extrapolation based on \ppdmany{}---which gives the most precise predictions---for empirical parameters in Fig.~\ref{fig:NM_EOS}. It higlights the fact that an uncertainty band is obtained, which is key for rigorous error estimates at higher densities. A multivariate Gaussian approximation of our \PPD{} for nuclear-matter parameters is provided as Supplemental Material~\cite{supp2022:PRL}. The interaction samples with importance weights are provided in the companion paper~\cite{jiang2022:long}.

{\it Summary and outlook---}%
In this work we constructed emulators that accurately reproduce full-space coupled-cluster computations of nucleonic matter starting from $\Delta$-full \chiEFT{} at \NNLO{} while reducing the computational cost by several orders of magnitude. The small-batch voting algorithm was developed to remove spurious states that occured in the \SPCC{} method. Using these tools, together with emulators of light nuclei, we employed history matching to identify more than one million acceptable interaction samples, and could reveal correlations between different properties of infinite nuclear matter. 
We then employed importance resampling to obtain \PPD[s], and studied the sensitivity to the choice of calibration data. Broad, asymmetric marginal distributions for the saturation energy and density were observed when making predictions conditional on few-nucleon data only, while the predictions become more precise and accurate when adding the \nuc{16}{O} energy and radius to the calibration data set.
Binding energies and radii of finite nuclei are obviously useful for model calibration, but are not the only choice. Other observables such as the \nuc{3}{H} beta-decay rate and three-nucleon scattering cross sections should also be considered within a statistical framework involving many-body predictions. Ongoing work shows that also these observables can be efficiently computed~\cite{Miller:2022beg, Miller:2022cil} or emulated~\cite{wesolowski2021, Zhang:2021jmi}.
Our rigorous error bands for the nuclear matter \EOS{} around saturation will be important for advances in studies of dense, neutron-rich matter and for the interpretation of nascent observations from multimessenger astonomy~\cite{Dietrich:2020efo,Huth:2021bsp,Lattimer:2023rpe}.

The introduction of small-batch voting demonstrates how emulators can be successfully constructed for challenging many-body observables. Further developments within various many-body computational frameworks are expected to follow~\cite{Duguet:2023wuh} in both nuclear physics and beyond. 
Future work using history matching, Bayesian \PPD{} sampling, and many-body emulators, as exemplified by this study, will help to elucidate the information content of various low-energy observables, the order-by-order convergence of \chiEFT{}, and the predictive power of \emph{ab initio} modeling across the nuclear landscape.

\begin{acknowledgments}
  We thank Andreas Ekstr\"om and Thomas Papenbrock for useful discussions. This work was supported by the Swedish Research Council (Grant Nos 2017-04234 and 2021-04507), the European Research Council under the European Unions Horizon 2020 research and innovation program (Grant No. 758027), and the U.S. Department of Energy under contract DE-AC05-00OR22725 with UT-Battelle, LLC (Oak Ridge National Laboratory).  
  The computations and data handling were enabled by resources provided by the Swedish National Infrastructure for Computing (SNIC) at Chalmers Centre for Computational Science and Engineering (C3SE), and the National Supercomputer Centre (NSC) partially funded by the Swedish Research Council through Grant No. 2018-05973.
\end{acknowledgments}

\bibliography{./master,./temp}
\end{document}

%% file: shorthands.tex
\newcommand{\prCond}[2]{\ensuremath{\mathrm{pr}\left(#1 \, \vert \, #2 \right)}}
\newcommand{\likelihood}[2]{\ensuremath{\mathcal{L}\left(#1 \, \vert \, #2 \right)}}
\newcommand{\Dcal}{\ensuremath{\mathcal{D}_\mathrm{cal}}}
\newcommand{\Dfew}{\ensuremath{\mathcal{D}_{A=2,3,4}}}
\newcommand{\Dmany}{\ensuremath{\mathcal{D}_{A=2,3,4,16}}}
\newcommand{\ppdfew}{\ensuremath{{\rm PPD}_{A=2,3,4}}}
\newcommand{\ppdmany}{\ensuremath{{\rm PPD}_{A=2,3,4,16}}}

\def\nuc#1#2{\relax\ifmmode{}^{#1}{\protect\text{#2}}\else${}^{#1}$#2\fi}
\def\itnuc#1#2{\setbox\@tempboxa=\hbox{\scriptsize\it #1}
\def\@tempa{{}^{\box\@tempboxa}\!\protect\text{\it #2}}\relax
\ifmmode \@tempa \else $\@tempa$\fi}

\newcommand{\newabbreviation}[3]{\newcounter{#1}\expandafter\newcommand\csname#1\endcsname[1][]{\ifthenelse{\equal{##1}{abreviate}}{#2}{\ifthenelse{\equal{##1}{fullname}}{#3}{\ifthenelse{\equal{##1}{explain}}{#3 (#2)\stepcounter{#1}}{\ifthenelse{\value{#1}=0}{#3##1 (#2##1)\stepcounter{#1}}{#2##1}}}}}}

\newabbreviation{EFT}{EFT}{effective field theory}
\newabbreviation{chiEFT}{$\chi$EFT}{chiral effective field theory}
\newabbreviation{LO}{\text{LO}}{leading order}
\newabbreviation{NLO}{\text{NLO}}{next-to-leading order}
\newabbreviation{NNLO}{\text{NNLO}}{next-to-next-to-leading order}
\newabbreviation{LEC}{\text{LEC}}{low-energy constant}

\newabbreviation{EC}{\text{EC}}{eigenvector continuation}
\newabbreviation{CC}{\text{CC}}{coupled-cluster}
\newabbreviation{NCSM}{\text{NCSM}}{no-core shell model}
\newabbreviation{SPCC}{\text{SPCC}}{subspace-projected coupled cluster}
\newabbreviation{PPD}{\text{PPD}}{posterior predictive distribution}
\newabbreviation{PDF}{\text{PDF}}{probability density function}
\newabbreviation{GP}{\text{GP}}{Gaussian processe}
\newabbreviation{NM}{\text{NM}}{nuclear matter}
\newabbreviation{PNM}{\text{PNM}}{pure neutron matter}
\newabbreviation{SNM}{\text{SNM}}{symmetric nuclear matter}
\newabbreviation{EOS}{\text{EOS}}{equation-of-state}

\newabbreviation{NN}{\text{NN}}{nucleon-nucleon}
\newabbreviation{TNF}{\text{3NF}}{three-nucleon force}

